\documentclass[
twocolumn,
superscriptaddress,
showpacs,
preprintnumbers,
amsmath,
amssymb,
aps,
pra,
]{revtex4-1}

\usepackage{graphicx}
\usepackage{dcolumn}
\usepackage{bm}
\usepackage{epstopdf}
\usepackage{tabularx}
\usepackage{array}
\usepackage{etoolbox}

\newcommand{\ket}[1]{\left|#1\right>} 
\newcommand{\bra}[1]{\left<#1\right|} 
\newcommand{\braket}[2]{\left<#1\vphantom{#2}|#2\vphantom{#1}\right>}
\newcommand{\ze}[1]{\,\mathrm{#1}}
\newcommand{\ti}[2]{#1_{\mathrm{#2}}}
\newcommand{\tio}[1]{_{\mathrm{#1}}}

\usepackage{siunitx}
\usepackage{color}
\usepackage{pgf}
\usepackage{etex}
\usepackage{pgfplots}
\usepackage{pgfplotstable}
\usepackage{fancybox}
\usetikzlibrary{backgrounds}
\usetikzlibrary{calc}
\pgfplotsset{width=7cm,compat=newest}

\usepackage{caption}
\usepackage{subcaption}
\usepackage[section]{placeins}

\usepackage[space]{grffile}
\usepackage{nicefrac}

\usepackage[pdftex,
       bookmarks=true,%
       bookmarksopen=true,
       bookmarksopenlevel=0,
       bookmarksnumbered=false,
       plainpages=false,
       hyperindex=true,
       allbordercolors=white,
       pdfstartview=,
       colorlinks=true,
       linkcolor=blue,
       allcolors=blue
 ]{hyperref}

 \hypersetup{%
    bookmarksdepth=2, bookmarksopen=true, bookmarksopenlevel=1,
}

\usepackage[capitalize]{cleveref}
\usepackage{bookmark}

\graphicspath{{figures_paper/}}

\begin{document}


\title[test]{Simultaneous sub-Doppler laser cooling of fermionic $^{\mathbf{6}}$Li and $^{\mathbf{40}}$K on the D$_{\mathbf{1}}$ line:\\
 Theory and Experiment}

\author{Franz Sievers}
\email{franz.sievers@lkb.ens.fr}
\affiliation{Laboratoire Kastler Brossel, \'Ecole Normale Sup\'erieure, CNRS, UPMC, 24 rue Lhomond, 75005 Paris, France}

\author{Norman Kretzschmar}
\affiliation{Laboratoire Kastler Brossel, \'Ecole Normale Sup\'erieure, CNRS, UPMC, 24 rue Lhomond, 75005 Paris, France}

\author{Diogo Rio Fernandes}
\affiliation{Laboratoire Kastler Brossel, \'Ecole Normale Sup\'erieure, CNRS, UPMC, 24 rue Lhomond, 75005 Paris, France}

\author{Daniel Suchet}
\affiliation{Laboratoire Kastler Brossel, \'Ecole Normale Sup\'erieure, CNRS, UPMC, 24 rue Lhomond, 75005 Paris, France}

\author{Michael Rabinovic}
\affiliation{Laboratoire Kastler Brossel, \'Ecole Normale Sup\'erieure, CNRS, UPMC, 24 rue Lhomond, 75005 Paris, France}

\author{Saijun Wu}
\email{saijunwu@fudan.edu.cn}
\affiliation{State Key Laboratory of Surface Physics and Department of Physics, Fudan University, 200433 Shanghai, P. R. China}

\author{Colin V. Parker}
\affiliation{James Franck Institute, Enrico Fermi Institute and Department of Physics,\\ University of Chicago, Chicago, IL 60637, USA}

\author{Lev Khaykovich}
\affiliation{Department of Physics, Bar-Ilan University, Ramat-Gan 52900, Israel}

\author{Christophe Salomon}
\affiliation{Laboratoire Kastler Brossel, \'Ecole Normale Sup\'erieure, CNRS, UPMC, 24 rue Lhomond, 75005 Paris, France}

\author{Fr\'ed\'eric Chevy}
\affiliation{Laboratoire Kastler Brossel, \'Ecole Normale Sup\'erieure, CNRS, UPMC, 24 rue Lhomond, 75005 Paris, France}

\date{\today}

\begin{abstract}
We report on simultaneous sub-Doppler laser cooling of fermionic $^6$Li and $^{40}$K using the D$_1$ optical transitions. We compare  experimental results to a numerical simulation of the cooling process applying a semi-classical Monte Carlo wavefunction method. The simulation takes into account the three dimensional optical molasses setup and the dipole interaction between atoms and the bichromatic light field driving the D$_1$ transitions. We discuss the physical mechanisms at play, we identify the important role of coherences between the ground state hyperfine levels and compare D$_1$ and D$_2$ sub-Doppler cooling. In 5\,ms, the D$_1$ molasses phase largely reduces the temperature for both $^6$Li and $^{40}$K at the same time, with a final temperature of 44\,$\mu$K and 11\,$\mu$K, respectively. For both species this leads to a phase-space density close to $10^{-4}$. These conditions are well suited to directly load an optical or magnetic trap for efficient evaporative cooling to quantum degeneracy.
\end{abstract}

\pacs{37.10.De, 32.80.Wr, 67.85.-d}

\bookmarksetup{depth=-1}

\maketitle

\bookmarksetup{depth=3}

\bookmarksetup{startatroot}


\newpage

\section*{\label{sec:Introduction}Introduction}

The road towards quantum degeneracy in atomic gases usually starts with a laser cooling and trapping phase. The resulting initial phase-space density of the atomic ensemble and the initial collision rate should be as large as possible for initiating efficient evaporative cooling to quantum degeneracy. Sub-Doppler cooling has proven to be a powerful technique to increase the phase-space density of most alkali atoms and other atoms with multiple level structure~\cite{Dalibard1989, Lett1989, Weiss1989}. However, in the case of lithium and potassium, the narrow excited-state structure of the D$_2$ transition compromises the efficiency of this cooling scheme~\cite{Bambini1997, Xu2003}. Both species possess stable fermionic and bosonic isotopes, and they play an important role in recent experimental studies of strongly correlated quantum gases. Thus, important efforts have been devoted to search for alternative laser cooling schemes.

For instance, it has recently been shown that three dimensional Sisyphus cooling for $^{7}$Li, some GHz red detuned from the D$_2$ line, can lead to temperatures as low as $100\ze{\mu K}$ with up to $45\ze{\%}$ of the atoms in the cooled fraction~\cite{Hamilton2014}. A second option is  to operate the magneto-optical trap (MOT) on a transition with smaller linewidth to reduce the Doppler temperature~\cite{Duarte2011, McKay2011, Sebastian2014}. Such transitions exist for $^6$Li and $^{40}$K in the near-UV and blue regions of the spectrum, respectively, leading to temperatures of $33\ze{\mu K}$ for $^6$Li and $63\ze{\mu K}$ for $^{40}$K. Yet, special optics and a coherent source at $323\ze{nm}$ for $^6$Li and $405\ze{nm}$ for $^{40}$K are needed for this approach. Additionally, at these wavelengthes the available power is still a limiting factor.

More recently a simpler sub-Doppler cooling scheme using blue detuned molasses operating on the D$_1$ line was proposed and demonstrated on $^{40}$K \cite{Fernandes2012} and has been extended to other atomic species such as $^{7}$Li \cite{Grier2013}, $^{39}$K \cite{Salomon2013, Nath2013b} and $^6$Li \cite{Burchianti2014a}. Using this technique, temperatures as low as 20~$\mu$K  ($^{40}$K), 50~$\mu$K  ($^{7}$Li), 6~$\mu$K  ($^{39}$K) and 40~$\mu$K  ($^{6}$Li) were reached.

Even though the main ingredients of the D$_1$ cooling scheme are now understood at a qualitative level, in particular the role of the coherences between hyperfine ground-state levels \cite{Grier2013}, a complete picture, taking into account the full level-structure of the atoms, is still missing. In this paper, we present a three-dimensional semi-classical solution of the optical Bloch equations that takes into account the full set of relevant energy levels of alkali atoms and we apply it to the case of $^6$Li and $^{40}$K. The model fully confirms the experimentally observed cooling behaviour and its robustness with respect to changes in experimental parameters. The model is validated by a good match between the simulation and the experimentally measured temperature and fluorescence rate.  We recover the important role of the Raman-detuning between the main cooling laser and the repumping laser on the achievable temperature. We show here for both $^6$Li and $^{40}$K, that the gain in temperature of a factor of $\sim$3 at the exact Raman-resonance is well reproduced by the theoretical model and that the amount of coherence between both hyperfine states shows a pronounced resonance behavior. Beyond individual studies of the two species, we also show experimentally that simultaneous cooling of $^6$Li and $^{40}$K does not lead to any severe trade-off and is technically easy to implement. We are able to capture more than $1\times 10^9$ atoms of each species, with a capture efficiency exceeding $60\ze{\%}$ from a compressed magneto-optical trap (CMOT), and reach temperatures as low as $44\ze{\mu K}$ for  $^6$Li and $11\ze{\mu K}$ for $^{40}$K within $5\ze{ms}$.

\section{D$_1$ cooling mechanism}

\begin{figure*}[htbp]
\begin{center}
\includegraphics[width=13cm]{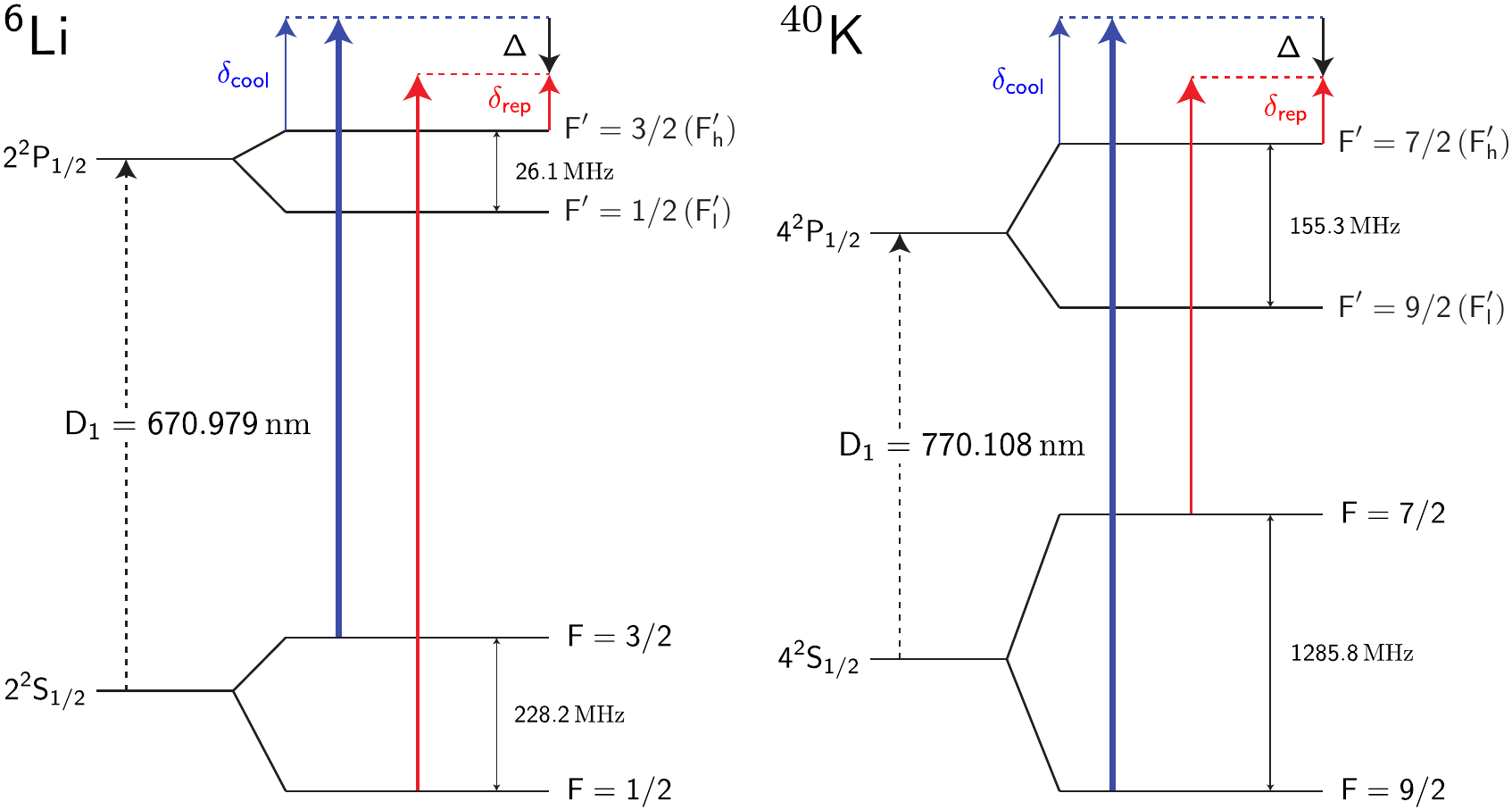}
\caption{\label{fig:Li_K_cooling_scheme}(Color online) Cooling scheme on the $^{6}$Li and $^{40}$K D$_1$ lines. The cooling beam (blue) is blue detuned by $\ti{\delta}{cool}$ from the $\ket{F=3/2}\rightarrow \ket{F_h'=3/2}$ ($\ket{F=9/2}\rightarrow \ket{F_h'=7/2}$) transition where $F'_h$ ($F'_l$) is the upper (lower) excited state level. The repumping beam (red) is blue detuned by $\ti{\delta}{rep}$ from the $\ket{F=1/2}\rightarrow \ket{F_h'=3/2}$ ($\ket{F=7/2}\rightarrow \ket{F_h'=7/2}$) transition. The detuning from the Raman-condition is denoted by $\Delta=\ti{\delta}{rep}-\ti{\delta}{cool}$.
}
\end{center}
\end{figure*}

In a typical D$_1$ cooling setup (\cref{fig:Li_K_cooling_scheme}), all the D$_1$ hyperfine levels are involved in the interaction. The sub-Doppler cooling effects include a mix of Sisyphus cooling, motion-induced and off-resonant light coupling from gray to bright levels, and coherent population trapping of slow atoms in nearly decoupled states. In this section we first introduce our semi-classical laser cooling model. We then present and compare the results from experimental observations and numerical simulations, and finally discuss the physical mechanism of D$_1$ cooling.

\subsection{Semi-classical Monte Carlo simulation}

The level diagrams of our bichromatic cooling scheme for both $^6$Li and $^{40}$K are depicted in \cref{fig:Li_K_cooling_scheme}. The D$_1$ molasses is composed of a 3D lattice whose polarization configuration is the same as that of a six-beam standard MOT, but with two sidebands to address the $\ket{F=3/2}$ and $\ket{F=1/2}$ hyperfine ground states of $^6$Li (resp. $\ket{F=9/2}$ and $\ket{F=7/2}$ for $^{40}$K) in the D$_1$ $\Lambda$-system at positive detunings.

Here, by convention, we refer to the $\ket{F=3/2}\rightarrow\ket{F'_h}$ and $\ket{F=1/2}\rightarrow\ket{F'_h}$ transitions as cooling/repumping transitions. It is however important to notice that neither the cooling nor the repumping D$_1$ transitions are actually closed.

Our numerical simulation of the cooling process is based on a semi-classical Monte Carlo wavefunction method. The simulation takes into account the three dimensional optical molasses setup and the dipole interaction between the single atoms and the polarized light driving the transitions of the D$_1$ manifold, which is spanned by the 4(2$I$+1) hyperfine Zeeman sub-levels ($I>0$ is the nuclear spin). We treat the external states of the atom classically and update its position ${\bf r}(t)$ and velocity ${\bf v}(t)$ according to the calculated expectation value of the light force
\begin{equation}
{\bf f}(t)=\frac{\bra{\psi(t)}-{\bf \nabla} H_{\rm eff}({\bf r}(t)) \ket{\psi(t)}}{\braket{\psi(t)}{\psi(t)}}.\label{eqf1}
\end{equation}
The atomic internal states $\ket{\psi(t)}$ evolve in a dressed basis with respect to the cooling laser (\cref{fig:Li_K_cooling_scheme}), according to the Monte Carlo wave function method~\cite{Dalibard1992,Carmichael1993} with the effective rotating-wave Hamiltonian
\begin{equation}
 H_{\rm eff}=H_0+H_{F=I-1/2}+H_{F=I+1/2}-i \hat \Gamma/2,
\end{equation}
where
\begin{align}
H_0 = & \sum_m \ket{F=I-1/2,m} \hbar \Delta \bra{F=I-1/2,m} \nonumber \\
& - \sum_{F' m'} \ket{F',m'} \hbar (\ti{\delta}{cool} +\delta_{{\rm hfs}, F'})\bra{F',m'}.
\end{align}
Here $H_0$ operates over the whole D$_1$ manifold, $\ti{\delta}{cool}$ is the detuning of the cooling laser with respect to the $F=I+1/2 \rightarrow F_h'$ transition, where $F_h'$ (\cref{fig:Li_K_cooling_scheme}) corresponds to the excited hyperfine level that is higher in energy, e.g. $F_h'=3/2$ for $^6$Li and $F_h'=7/2$ for $^{40}$K. $\Delta=\ti{\delta}{rep}-\ti{\delta}{cool}$ is the two-photon detuning for the $F=I-1/2 \rightarrow F=I+1/2$ Raman-transition, and $\delta_{{\rm hfs},F'}$ the hyperfine splitting of the excited state for $F'_l$ and zero for $F'_h$. The zero of energy is chosen as that of the bare F=I+1/2 ground state.

The light-atom coupling Hamiltonian
\begin{align}
H_{F=I\pm 1/2}=\hbar \sum_{m,\sigma,F',m'} & \Omega_{F,\sigma} ~c_{F,m,\sigma,F',m'} \nonumber\\
& \times\ket{F,m} \bra{F',m'}+h.c.
\end{align}
describes the cooling ($F=I+1/2$) and repumping ($F=I-1/2$) interactions\,\footnote{By assuming the ground-state hyperfine splitting to be large enough, the $F=I+1/2$ ($F=I-1/2$) is dark to the repumping (cooling) laser and the associated couplings are ignored in the Hamiltonian. This approximation allows us to write down the Hamiltonian in the time-independent form.}. Here $\Omega_{F,\sigma}$ are the Rabi frequencies of the repumping and cooling laser beams for $F=I-1/2,\,I+1/2$ respectively.
$c_{F,m,\sigma,F' m'}$ represent the Clebsch–-Gordan coefficients associated with the transitions coupled by $\sigma$ polarized light. To take into account the radiation damping we include the spontaneous emission rate $\hat \Gamma=\Gamma \hat P_{ee}$ where $\ze{\Gamma}$ is the excited-state linewidth and $\hat P_{ee}=\sum_{F',m'}\ket{F'm'} \bra{F'm'}$. This leads to a decay of the internal state wave function norm $\langle\psi(t)\ket{\psi(t)}$. The speed of this decay probabilistically dictates the collapse of the internal quantum states in the numerical simulation, which corresponds to spontaneous emission. We take into account the polarization of the spontaneous scattering photon and follow the standard quantum jump procedure to project the atomic states to ground states with its norm reset to unity~\cite{Dalibard1992}. To effectively account for heating due to both absorption and spontaneous emission, we assign a recoil momentum shift to ${\bf v}(t)$ twice before continuing to evolve $\ket{\psi(t)}$ via $H_{\rm eff}({\bf r}(t))$.

The simulations are performed with parameters matching the experimental setup by properly introducing the spatially-dependent $\Omega_{F,\sigma}({\bf r})$, the detunings $\Delta$, $\delta_{\rm cool}$, and atomic initial conditions. To reproduce experimental conditions, we fix the relative phases for all the 12 cooling and repumping laser beams at values randomized for each simulation trial. We record the evolution of the 3D atomic velocity,  the time-stamped fluorescence events corresponding to quantum jumps, as well as internal states properties such as state population and coherence. The observables are averaged over multiple simulation trials for comparison with the experiment.

\subsection{Raman-detuning dependence for $^6$Li}

\begin{figure*}
        \centering
        \begin{subfigure}[b]{8cm}
                \centering
                \resizebox{.976\linewidth}{!}{
                \includegraphics[width=\linewidth]{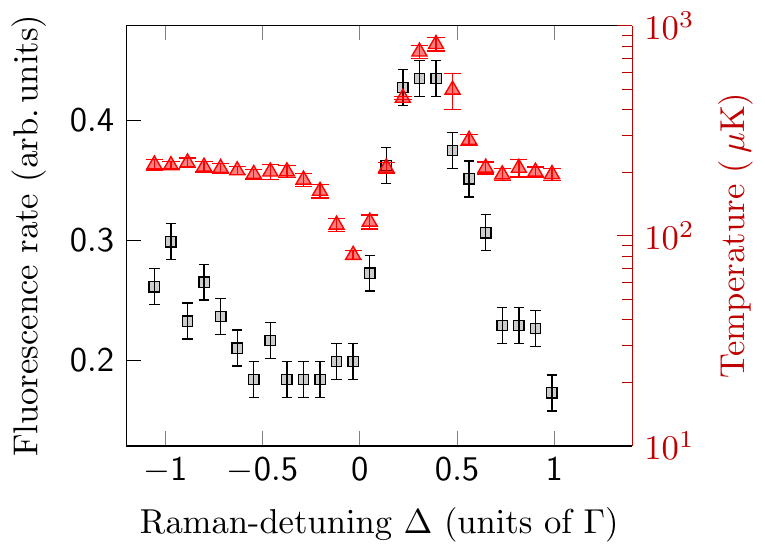}
                }
                \caption{Experiment: Low intensity}
                \label{fig:Li_rep_detuning_low_exp}
        \end{subfigure}%
        ~
        \begin{subfigure}[b]{8cm}
                \centering
                \resizebox{.976\linewidth}{!}{
                \includegraphics[width=\linewidth]{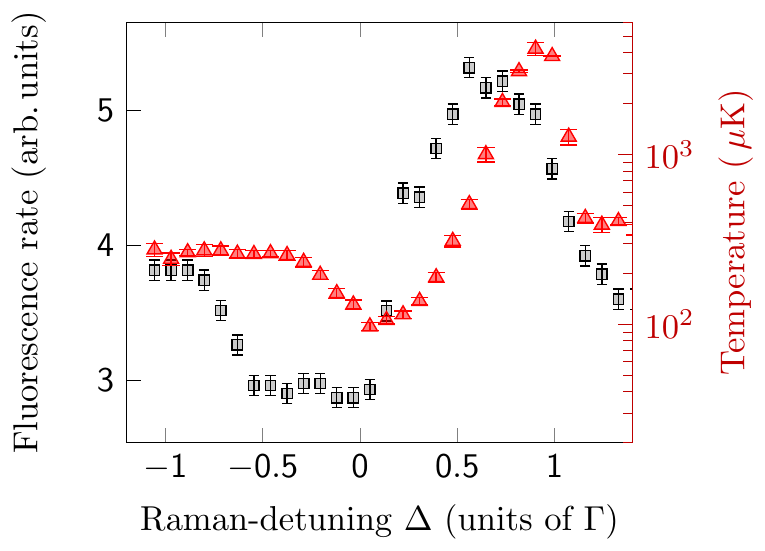}
                }
                \caption{Experiment: High intensity}
                \label{fig:Li_rep_detuning_high_exp}
        \end{subfigure}\\ \vspace{5 mm}
        \centering
        \begin{subfigure}[b]{8cm}
                \centering
                \resizebox{.976\linewidth}{!}{
                \includegraphics{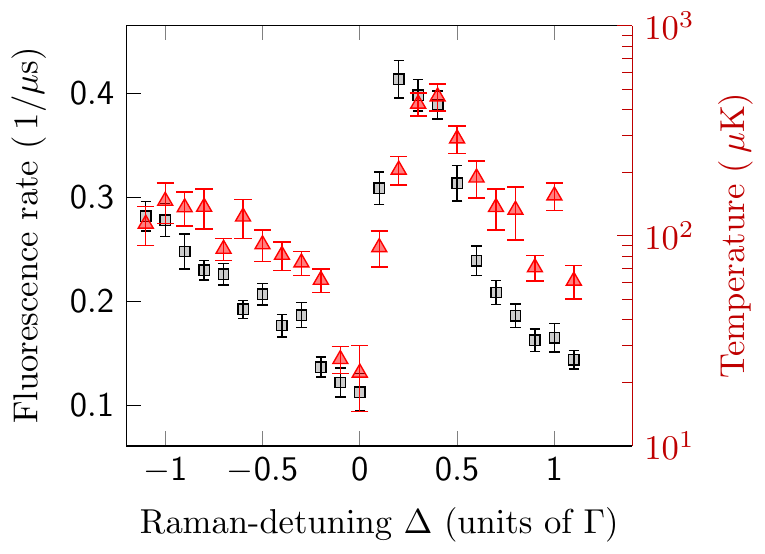}
                }
                \caption{Simulation: Low intensity}
                \label{fig:Li_rep_detuning_low_sim}
        \end{subfigure}%
        ~
        \begin{subfigure}[b]{8cm}
                \centering
                \resizebox{.976\linewidth}{!}{

                \includegraphics{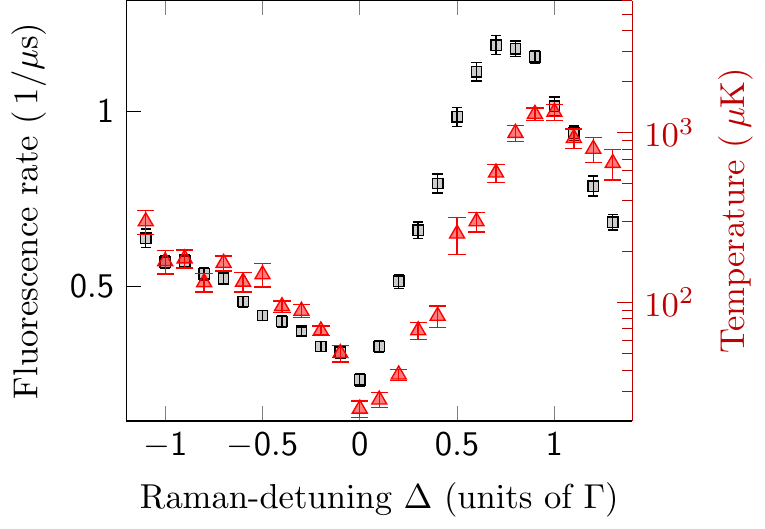}
                }
                \caption{Simulation: High intensity}

                \label{fig:Li_rep_detuning_high_sim}
        \end{subfigure}

        \caption{(Color online) Fluorescence (squares) and temperature (triangles, logarithmic scale) of the $^6$Li atomic cloud after a $100\ze{\mu s}$ pulse of D$_1$ light with variable Raman-detuning $\Delta$. (\protect\subref{fig:Li_rep_detuning_low_exp}) and (\protect\subref{fig:Li_rep_detuning_low_sim}) show the experimental and simulation results for $\ti{I}{cool}=2.7\,{\ti{I}{sat}}$, $\ti{I}{rep}=0.13\,{\ti{I}{sat}}$, (\protect\subref{fig:Li_rep_detuning_high_exp}) and (\protect\subref{fig:Li_rep_detuning_high_sim}) for $\ti{I}{cool}=9\,{\ti{I}{sat}}$, $\ti{I}{rep}=0.46 \,{\ti{I}{sat}}$ per beam.}\label{fig:Li_fluorescence}
\end{figure*}
\begin{figure}
\centering
        \begin{subfigure}[]{8cm}
                \includegraphics{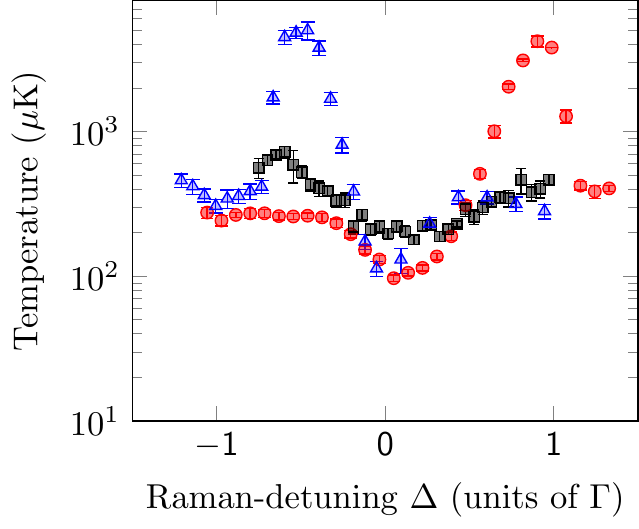}
                \caption{Experiment}
                \label{fig:Li_rep_detuning_intensity-ratio_exp}
        \end{subfigure}
        \centering
        \begin{subfigure}[]{8cm}
                \includegraphics{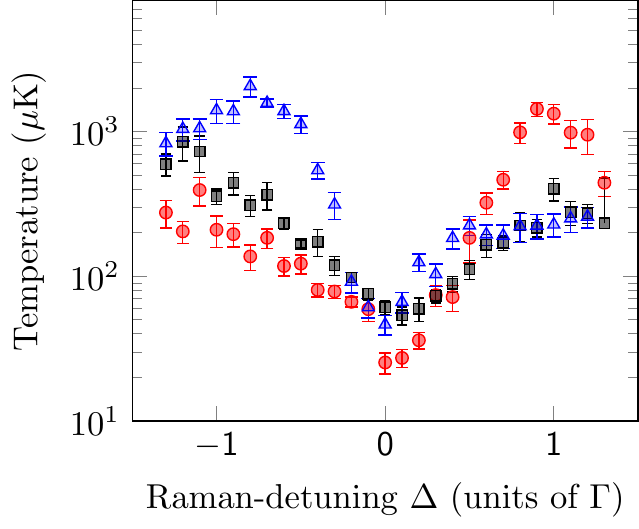}
                \caption{Simulation}
                \label{fig:Li_rep_detuning_intensity-ratio_sim}
        \end{subfigure}
        \caption{\label{fig:Li_rep_detuning_intensity-ratio_exp_and_sim}(Color online) Temperature of the $^6$Li D$_1$ molasses after a $100\ze{\mu s}$ pulse with variable Raman-detuning $\Delta$ for different cooling and repumping intensities. Standard intensities (red circles): $\ti{I}{cool}=9\,{\ti{I}{sat}}$, $\ti{I}{rep}=0.46 \,{\ti{I}{sat}}$. Equal cooling/repumping ratio (black squares): $\ti{I}{cool}=\ti{I}{rep}=9\,{\ti{I}{sat}}$. Inverted cooling/repumping ratio (blue triangles): $\ti{I}{cool}=0.6\,{\ti{I}{sat}}$, $\ti{I}{rep}=4.6 \,{\ti{I}{sat}}$.}
\end{figure}
A critical parameter in the D$_1$ molasses scheme is the Raman-detuning $\Delta$ (\cref{fig:Li_K_cooling_scheme}). In the following we investigate the Raman-detuning dependence of the $^6$Li molasses temperature and fluorescence rate both theoretically and experimentally, for various cooling and repumping laser intensities.

Our $^6$Li-$^{40}$K machine is described in \cite{Ridinger2011}. We first load a lithium MOT using a laser slowed atomic beam (Zeeman slower). After a compressed MOT phase the magnetic field and the D$_2$ light are switched off and the D$_1$ molasses is applied (a more detailed description of the sequence is presented in the \hyperref[sec:Experimental_Setup]{Appendix}).
To probe the Raman-detuning dependence we apply a $100\ze{\mu s}$ D$_1$ molasses pulse with variable $\Delta$ to an atomic cloud precooled to $100\ze{\mu K}$.
\Cref{fig:Li_rep_detuning_low_exp,fig:Li_rep_detuning_high_exp} show the fluorescence rate and the temperature after the pulse as functions of the Raman-detuning $\Delta$ for the intensities used in the simulations.
We observe a temperature dip at zero Raman-detuning and a heating/fluorescence peak at positive $\Delta$ whose position and amplitude are correlated to the molasses intensity.

In the simulations we set the initial velocity of lithium to 0.2~m/s ($T\sim$ 30~$\mu$K). The simulation time is set to 200~$\mu$s. In the first $100~\mu$s we allow the cooling dynamics to equilibrate, and during the second $100~\mu$s  we record the velocity ${\bf v}(t)$  as well as the time-stamped quantum jump events  to calculate the equilibrium temperature and fluorescence rate. At each Raman-detuning we average over 25 trajectories.
The simulation results for two different intensities $\ti{I}{cool}=2.7\,{\ti{I}{sat}}$, $\ti{I}{rep}=0.13\,{\ti{I}{sat}}$ and $\ti{I}{cool}=9\,{\ti{I}{sat}}$, $\ti{I}{rep}=0.46 \,{\ti{I}{sat}}$ are shown in \cref{fig:Li_rep_detuning_low_sim,fig:Li_rep_detuning_high_sim}, respectively (here $I_{\rm sat}$ refers to the saturation intensity of the D$_2$ line).

The simulated heating/fluorescence peak positions for low and high intensities (\cref{fig:Li_fluorescence}) agree well with the experimental findings. Also the shift between the heating and fluorescence peak, which increases with the molasses intensity, is numerically reproduced without any freely adjustable parameters.

Despite the nice match between simulations and experiments in \cref{fig:Li_fluorescence,fig:Li_rep_detuning_intensity-ratio_exp_and_sim}, we observe that the semi-classical simulations provide temperatures that are systematically a factor two to four lower than the measured ones, particularly near the Raman-resonance condition $\Delta=0$. Here the simulation predicts a temperature of 20~$\mu$K whereas the lowest measured temperature is 50~$\mu$K. The reason for this is not fully understood and may come both from theory and experimental limitations. First the simulation is semi-classical and neglects the wavefunction extent of the cold atoms. The predicted temperature of 20\,$\mu$K  corresponds to only six times the recoil energy $E_R= \frac{1}{2}m v^2_{\rm recoil}=k_B\times3.5 \mu$K. Therefore, only a quantum treatment of the atom’s external motion can be expected to give a quantitative equilibrium temperature prediction in the low intensity limit. In the simulation we observe that slow atoms are likely trapped within sub-wavelength regions, where the light shift is minimal and the atom is nearly decoupled from light over a long time without quantum jump. This coherent population trapping effect enhances the cooling at both large and small $\Delta$, although it is most pronounced at the Raman-resonance ($\Delta=0$) since more choices of decoupled states emerge. The semi-classical picture clearly exaggerates the cooling effect since the wave nature of the atom's external motion is not included in the model. In fact, the wave function of the slow atoms will sample a larger volume of the sub-wavelength traps and will shorten the lifetime of the dark periods.

On the experimental side the residual magnetic field cancellation has only been coarsely tuned for the data set presented in \cref{fig:Li_fluorescence} (as well as in \cref{fig:Li_rep_detuning_intensity-ratio_exp_and_sim,fig:K_Rep_detuning_exp}). With careful tuning of the magnetic field zeroing we were able to lower the $^{40}$K temperature to 11\,$\mu$K (\cref{subsec:raman_K} on $^{40}$K) for lower density samples. Interestingly, other groups have indeed found on $^{39}$K lower temperatures (6\,$\mu$K) than ours at the Raman condition\,\cite{Salomon2013}. Note also that in \cref{fig:Li_fluorescence} for positive Raman detunings ($\Delta\sim 0.5~\Gamma$ at low intensity and $\Delta\sim\Gamma$ at high intensity) the ``temperature'' corresponds to out of equilibrium situations as the atoms are quickly heated away and lost from the molasses.  The notion of temperature should thus be taken with care in this region, unlike for negative Raman detunings where a steady-state temperature is reached.

Another reason for shortening the lifetime of dark periods of the slow atoms is re-absorption of photons emitted by other atoms. We have indeed seen a density dependent excess temperature which we measured to be $4.6\ze{\mu {\rm K}\times10^{11}~at./cm^3}$ for $^{40}$K. A careful simulation of cooling including photon re-absorption processes is far more complex and is beyond the scope of this work.

We also study the same Raman-detuning dependent effects, but for different cooling/repumping ratios. Typical experimental and simulation results are presented in \cref{fig:Li_rep_detuning_intensity-ratio_exp_and_sim}. Here again, the simulation parameters are chosen according to the experimental values. The simulation and experiments match fairly well. In particular, for the usual configuration with $\ti{I}{cool}/\ti{I}{rep}>1$ ($\ti{I}{cool}=9~\ti{I}{sat}$ and $\ti{I}{rep}=0.45~\ti{I}{sat}$), we observe a heating peak at $\Delta >0$.
When inverting the roles of the cooling and repumping light, i.e. $\ti{I}{cool}/\ti{I}{rep}<1$ ($\ti{I}{cool}=0.6~\ti{I}{sat}$ and $\ti{I}{rep}=4.6~\ti{I}{sat}$), the heating peak appears for $\Delta < 0$ instead. In all cases, cooling is  most efficient at the Raman-resonance ($\Delta=0$). Finally, for $\ti{I}{cool}$ equal to $\ti{I}{rep}$, both as large as $9~\ti{I}{sat}$, we observe less efficient cooling at $\Delta=0$ with moderate heating at blue and red detunings.

\subsection{Raman-detuning dependence for $^{40}$K
\label{subsec:raman_K}}

\begin{figure}
\begin{center}
\includegraphics{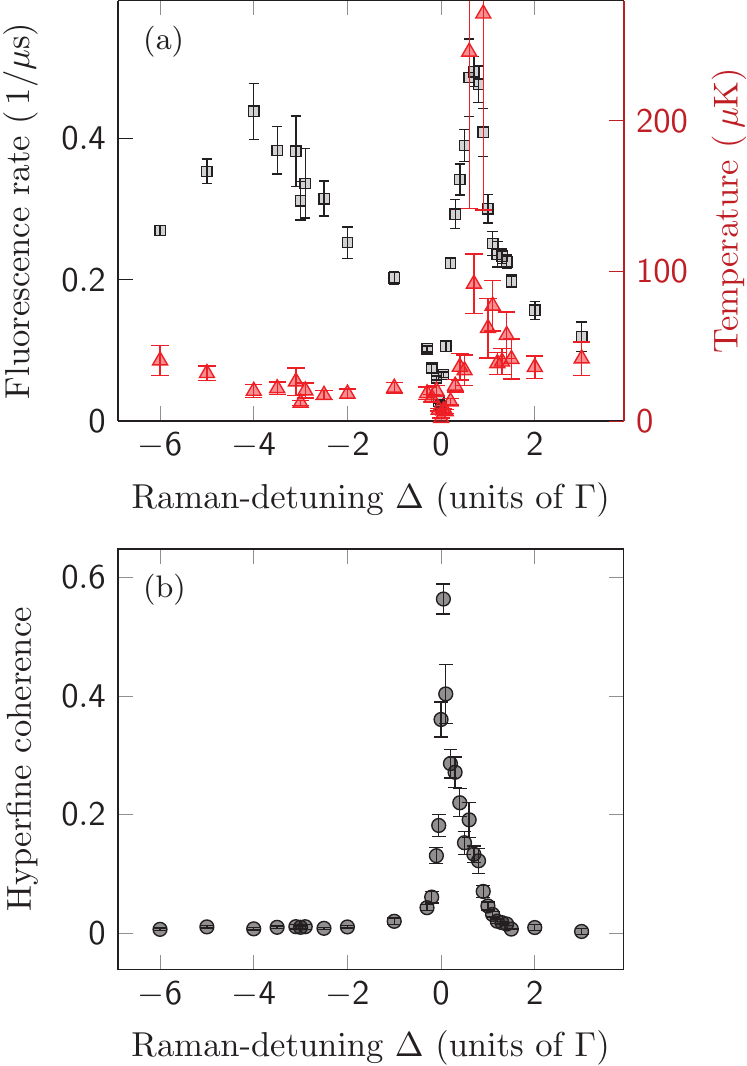}
\caption{\label{fig:K_coherence_sim}(Color online) Simulation: Hyperfine coherence and $\Lambda$-enhanced cooling for the $^{40}$K D$_1$ molasses. Simulation time 2\,ms, $\ti{\delta}{cool}=3\ze{\Gamma}$, $\ti{I}{cool}=6\,{\ti{I}{sat}}$, $\ti{I}{rep}/\ti{I}{cool}=7.6\ze{\%}$. (a) Fluorescence (squares) and temperature (triangles) as functions of the Raman-detuning $\Delta$. (b) Coherence $4\cdot\langle\rho_{F=7/2,F=9/2}^2\rangle$ between the two hyperfine ground states $F=7/2$ and $F=9/2$ (see \cref{subsec:Physical_picture}). The coherence is peaked at the Raman-resonance condition, with a width matching the temperature dip.}
\end{center}
\end{figure}

Typical simulation results for $^{40}$K are shown in \cref{fig:K_coherence_sim}a. Compared to $^6$Li, simulations for $^{40}$K require significant higher computational power due to larger internal state dimensions as well as a larger atomic mass and therefore slower cooling dynamics.
To save computation time, we start at a velocity of 0.2~m/s ($T\sim$ 50~$\mu$K), and set the simulation time to 2~ms. We record the velocity ${\bf v}(t)$ as well as the time-stamped quantum jump events for $t>1\ze{ms}$ to calculate the fluorescence rate. For each Raman-detuning $\Delta$, 13 trajectories are simulated.

\begin{figure}
        \begin{center}
        \includegraphics{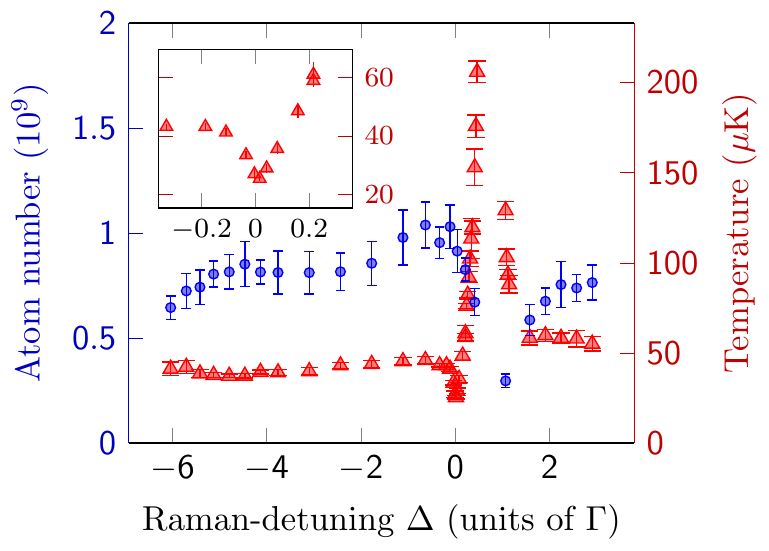}
        \end{center}
        \caption{\label{fig:K_Rep_detuning_exp}(Color online) Experiment: Atom number and equilibrium temperature of the $^{40}$K D$_1$ molasses as functions of the Raman-detuning $\Delta$. $\ti{\delta}{cool}=3\ze{\Gamma}$,
        $\ti{I}{cool}=6\,{\ti{I}{sat}}$,
        $\ti{I}{rep}/\ti{I}{cool}=7.6\ze{\%}$,
        $\ti{t}{m}=5\ze{ms}$. In the constant temperature regions below $-0.1\ze{\Gamma}$ and above $2\ze{\Gamma}$  gray molasses cooling involves  coherences between Zeeman states in a given hyperfine state but not between hyperfine states. At the exact Raman-condition $\Delta=0$, long-lived coherences between hyperfine states are established, as can be seen in the simulation in \cref{fig:K_coherence_sim}. In a narrow detuning range, the temperature (red triangles) drops to 20$\ze{\mu K}$ (inset: expanded scale).}
\end{figure}

Experimental results for $^{40}$K are presented in \cref{fig:K_Rep_detuning_exp}, showing the temperature and atom number of the D$_1$ molasses as functions of the Raman-detuning $\Delta$. The total molasses duration $\ti{t}{m}=5\ze{ms}$.  In the last 2 ms a linear intensity ramp to $\ti{I}{cool}=6\,{\ti{I}{sat}}$ is performed.
Just like $^6$Li, we observe a sharp temperature drop at the Raman-condition, a heating resonance at ${\sim}0.7\ze{\Gamma}$ and constant temperature regions below $-0.1\ze{\Gamma}$ and above $2\ze{\Gamma}$. For the constant temperature regions the temperature $T\sim45\ze{\mu K}$ is small compared to the Doppler-temperature $\ti{T}{Doppler,K}=145\ze{\mu K}$.
On the Raman-condition the temperature decreases to $23\ze{\mu K}$.  In carefully optimized conditions we measured temperatures as low as $11\ze{\mu K}$.

As for $^6$Li, the comparison between \cref{fig:K_Rep_detuning_exp} and \cref{fig:K_coherence_sim}a again demonstrates the qualitative agreement between simulations and experimental results and that the heating peak position is reproduced by the simulation without adjustable parameters. Interestingly the inverted hyperfine structure in the ground and excited states of $^{40}$K and the different $F\to {F'=F-1}$ transition for the cooling laser and $F\to {F'=F}$ repumping transition does not significantly modify the D$_1$ cooling scheme as compared to $^{6}$Li.
\begin{figure*}
    \begin{center}
    \includegraphics{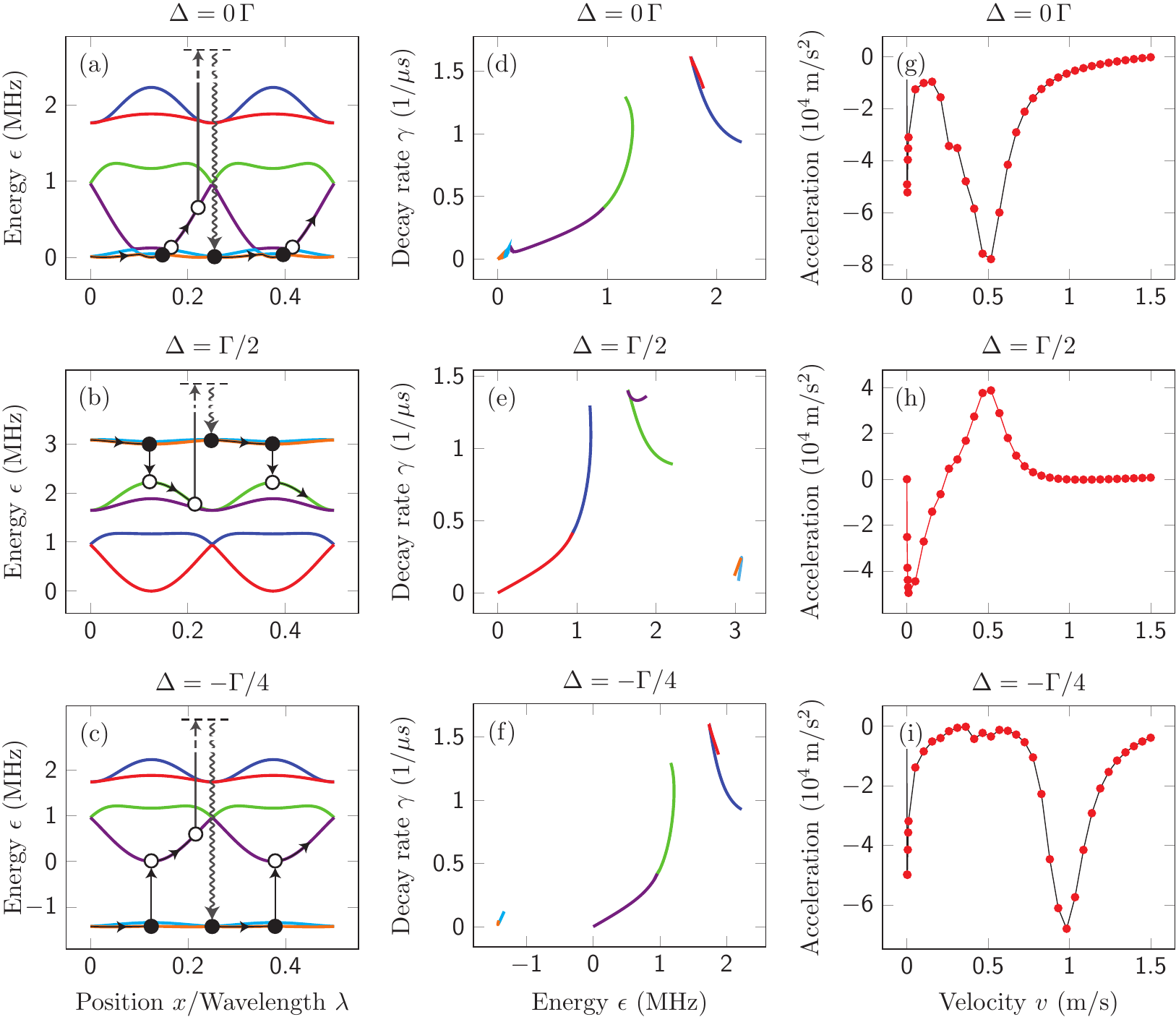}
    \caption{(Color online) Cooling mechanism around Raman-condition in a simplified model.  Optical Bloch equation simulation for $^6$Li subjected to a 1D  bichromatic lattice with linear orthogonal polarizations near D$_1$ resonance. The cooling lattice and repumping lattice are displaced by $\pi$ ($\lambda/4$). $\ti{I}{cool}=15\,{\ti{I}{sat}}$, $\ti{I}{rep}=0.75\,{\ti{I}{sat}}$, $\ti{\delta}{cool}=4\ze{\Gamma}$. (a)(b)(c) dressed states as functions of the position in units of the D$_1$ optical wavelength. The two dressed $F=1/2$ levels (light blue and orange) are nearly flat in all the graphs due to the small $\ti{I}{rep}$. (d)(e)(f) show the decay rate of dressed states as a function of their energy shifts. Here the two dressed $F=1/2$ levels span a very small energy range and are with low decay rate. (g)(h)(i) show the velocity-dependent optical force for an atom dragged with velocity $v$. Figs. (a)(d)(g) are for $\Delta=0$; Figs. (b)(e)(h) are for $\Delta=\Gamma/2$. Figs. (c)(f)(i) are for $\Delta=-\Gamma/4$. Note the negative sign of the force in (g) and (i) implying cooling, and the anti-cooling force for velocities near 0.5 m/s in (h).}\label{Fig:1D_OBE}
    \end{center}
\end{figure*}

\subsection{The D$_1$ cooling mechanism}

The agreement between simulation and experiment suggests that the semi-classical picture is able to catch the essential physics behind the D$_1$ molasses cooling. In particular, the  mechanisms behind the cooling dips and heating peaks in \cref{fig:Li_fluorescence,fig:Li_rep_detuning_intensity-ratio_exp_and_sim,fig:K_Rep_detuning_exp}, previously interpreted using the dressed atom picture with a simplified three-level model~\cite{Grier2013}, survive in the full level scheme of the D$_1$ transition.

It is well known that efficient D$_2$ sub-Doppler cooling requires isolated excited hyperfine levels for alkaline atoms~\cite{Bambini1997, Xu2003}. In contrast, the D$_1$ gray molasses operates well even when all  D$_1$ levels are excited (as is the case of $^6$Li) and even at zero excited-state hyperfine splitting as confirmed numerically. The robustness of D$_1$ molasses is also seen in its insensitivity to the relative phase between the ``cooling'' and ``repumping'' lattices, a critical parameter for D$_2$ bichromatic cooling where no polarization gradient was introduced~\cite{Gupta1993}.

In the following we discuss the physics behind the robustness of the D$_1$ sub-Doppler cooling. We then revisit the cooling dips and heating peaks in \cref{fig:Li_fluorescence,fig:Li_rep_detuning_intensity-ratio_exp_and_sim,fig:K_Rep_detuning_exp}.

We notice all the dipole allowed D$_1$ transitions (\cref{fig:Li_K_cooling_scheme}) are ``open'': when addressed with weak off-resonant light, the probability of inelastic ($m_F$- or $F$-changing) photon scattering is comparable to or larger than that of elastic scattering.  When blue detuned from the D$_1$ transitions, an off-resonant bichromatic lattice can establish a correlation between the spatially varying light shift (due to virtual elastic scattering) and decay (due to real inelastic scattering) for the dressed ground states, since a larger light shift is accompanied with a stronger light-atom coupling and typically a larger inelastic scattering cross-section.

We verify this idea with the full D$_1$ model for $^6$Li atoms subjected to a 1D lattice with orthogonal linear polarizations (lin$\perp$lin configuration) with typical cooling parameters. The spatially varying light shifts $\epsilon$ of the six dressed ground states of $^6$Li are plotted in \cref{Fig:1D_OBE}a. The decay of the dressed states, due to inelastic light scattering, are characterized by the decay rate $\gamma$ that is plotted versus $\epsilon$ in \cref{Fig:1D_OBE}d. We see a correlation between $\epsilon$ and $\gamma$ for $\epsilon<1.5$~MHz. Such correlation is robustly established for the D$_1$ transitions, as verified numerically in the more complicated 3D lattices and for other atomic species. The correlation even persists for a fictitious atom with vanishing D$_1$ hyperfine splitting and reduced $m_F$-changing light scattering~\cite{Happer1972}.

Such a correlation between the spatially dependent light shift $\epsilon$ and decay rate $\gamma$ has two consequences: First, atoms with $k v<\gamma$ tend to accumulate in dressed states with low light shifts, which facilitates cooling through motion-induced coupling to higher energy dressed states~\cite{Cohen-Tannoudji1990a}. This coupling is non-adiabatic as in the famous $\sigma^+$--$\sigma^-$ 1D velocity selective coherent population trapping sub-recoil cooling mechanism (VSCPT), where the spatial gradient of the atom’s wavefunction induces a coupling to a bright state.
Second, for a slowly moving atom that adiabatically follows a particular dressed state, the atom tends to leave the dressed state when the light shift is high, leading to Sisyphus cooling. In addition, at locations where $\epsilon,\gamma\sim 0$, slow atoms can be confined near the local dark states such as those in \cref{Fig:1D_OBE}a near $x=0,\,\lambda/8,\,\lambda/4,\,3\lambda/8$~\footnote{Exact locations of the dark states depend on the relative phase between the cooling and repumping lattices.}. The resulting optical cooling force is plotted in \cref{Fig:1D_OBE}g and is negative (cooling effect) over a broad range. We emphasize that this  simplified 1D analysis remains valid in the more complex 3D beam geometry and is not restricted  to $^6$Li atoms. The D$_1$ laser cooling mechanism applies to all alkalis even those amenable to efficient D$_2$ sub-Doppler cooling such as cesium or rubidium. As D$_1$ laser cooling involves dark states it is less affected by density dependent photon multiple scattering and heating than D$_2$ sub-Doppler cooling. Therefore it would be interesting to quantify the gain in phase-space density by applying D$_1$ sub-Doppler cooling for these atoms.

In comparison, sub-Doppler cooling on the D$_2$ lines is significantly different. While the $F=I\pm 1/2 \rightarrow F'=I\pm 1/2$ transitions are as ``open'' as in D$_1$, the $F=I+1/2 \rightarrow F'=I+3/2$ and $F=I-1/2 \rightarrow F'=I-3/2$ have both ``closed'' and ``open'' transitions. Here the ``closed'' transitions are characterized by a greater-than-unity elastic-to-inelastic scattering ratio. If the $F=I+1/2 \rightarrow F'=I+3/2$ transitions can be isolated, then by taking advantage of the nearly closed $m_F-m_{F'}$ transitions, a correlation between the light shift and decay rate can be established with (instead) a red detuned lattice, as in standard sub-Doppler cooling~\cite{Dalibard1989, Lett1989, Weiss1989}. However, in case of small hyperfine splitting the ``open'' hyperfine transitions are as well addressed at red detuning, leading to short-lived potential minima and degraded correlations, contributing to the inefficiency of the sub-Doppler cooling~\footnote{For the D$_2$ transitions the complication is also $\Delta$-dependent. A complete comparison of D$_1$ and D$_2$ cooling will be the subject of a future publication.}.

\subsection{Physical picture of the Raman-detuning effect
\label{subsec:Physical_picture}}

We now extend the three-level picture of Ref.~\cite{Grier2013} to understand the details of the experiment. The cooling dips observed both experimentally and numerically at the Raman-resonance condition are also fairly easy to understand in the full model: At $\Delta=0$ the resonant Raman-coupling splits the $F=I\pm1/2$ hyperfine ground states into a bright and a dark manifold. The dark manifold is weakly coupled to the molasses. More precisely, the coupling strength of the Raman dark manifold is even weaker than those due to individual cooling/repumping couplings. Therefore the emergence of the Raman dark manifold enhances all sub-Doppler cooling effects.

Since the dark manifold is a coherent superposition of the two hyperfine states $F_1=I-1/2$ and $F_2=I+1/2$, we expect that laser cooled atoms mostly occupy the dark manifold and  therefore display a hyperfine coherence $\rho_{F_1,F_2}$ with significant amplitude. To test this picture, we record the time-dependent off-diagonal density matrix quantity $4\cdot\langle \rho_{F_1,F_2}^2(t) \rangle$ for all the quantum trajectories of the numerical simulations. The factor 4 ensures the normalization to one for the maximally coherent situation. To compute the two-photon detuning $\Delta$ dependent quantity $4\cdot\langle \rho_{F_1,F_2}^2 \rangle$ we average over both the equilibrium time and many quantum trajectories at fixed $\Delta$. Typical results for $^{40}$K are given in \cref{fig:K_coherence_sim} with the cooling parameters corresponding to \cref{fig:K_Rep_detuning_exp}. We see that the coherence $4\cdot\langle \rho_{F_1=7/2,F_2=9/2}^2 \rangle$ is peaked at the Raman-resonance condition and becomes significant with a width matching the temperature dip.

As in \cref{fig:Li_fluorescence,fig:Li_rep_detuning_intensity-ratio_exp_and_sim,fig:K_Rep_detuning_exp} and~\cite{Grier2013} we now explain the heating peaks with the full D$_1$ model. We first focus on the case of $\ti{I}{cool}\gg\ti{I}{rep}$ so that at large $|\Delta|$ the dressed $F=I-1/2$ hyperfine level is relatively long-lived and populated. As in Figs.~\ref{Fig:1D_OBE}b,c, the Raman-detuning $\Delta$ determines the energy level of the dressed $F=I-1/2$ hyperfine level, and it is clear that when $\Delta<0$, the motion-induced-coupling to the dressed $F=I+1/2$ level should still contribute to cooling (as in Figs.~\ref{Fig:1D_OBE}c,f,i)~\cite{Cohen-Tannoudji1990a}, apart from Sisyphus cooling. On the contrary, for $\Delta>\max[\epsilon_{F=I+1/2}]$, e.g. a Raman-detuning beyond the maximum light shift of the dressed $F=I+1/2$ manifold (as in Fig.~\ref{Fig:1D_OBE}b,e,h), motion-induced coupling to the lower energy dressed $F=I+1/2$ manifold would lead to heating. In addition, the Sisyphus effect at the $F=I+1/2$ manifold also contributes to heating, since atoms coupled from $F=I-1/2$ are more likely to start at the anti-trap positions. The corresponding heating peak is located at $\Delta\sim\max[\epsilon_{F=I+1/2}]>0$.

When $\ti{I}{cool}\ll\ti{I}{rep}$, the heating peak is shifted to $\Delta\sim -\max[\epsilon_{F=I-1/2}]<0$, as in \cref{fig:Li_rep_detuning_intensity-ratio_exp_and_sim}. This is straightforward to understand as the role of the two hyperfine ground states are now inverted with respect to the previous case. Finally, for $\ti{I}{cool}\sim \ti{I}{rep}$, the two hyperfine ground states have similar lifetimes and therefore similar steady-state populations. As the heating effects are balanced by cooling effects, the corresponding heating peaks in \cref{fig:Li_rep_detuning_intensity-ratio_exp_and_sim} (black squares) at $\Delta\sim\max[\epsilon_{F=I+1/2}]$, $\Delta\sim-\max[\epsilon_{F=I-1/2}]$ are substantially suppressed.

\section{Simultaneous $^{6}$L\lowercase{i} and $^{40}$K D$_1$ cooling}

Finally, we discuss the simultaneous operation of the $^{6}$Li and $^{40}$K D$_1$ molasses. We found that this simultaneous operation is required for subsequent efficient thermalization between both species in a quadrupole magnetic trap. The timing sequence and parameters are the same as for single-species operation. Experimental details are given in the \hyperref[sec:Experimental_Setup]{Appendix} and in \cite{Fernandes2012}). The D$_1$ molasses phase is composed of a 3\,ms capture phase and a 2\,ms cooling phase. \Cref{tab:D1_Molasses_Parameters} summarizes the optimal parameters of the dual-species molasses.
%
\begin{table}[htbp]
\begin{minipage}{\columnwidth}
\centering
\begin{ruledtabular}
\renewcommand{\arraystretch}{1.2}
\begin{tabular}{lcc}
 & \textsc{Potassium} & \textsc{Lithium}\\
\colrule
$P$ (mW)                     & 230   & 300 \\
$\ti{\delta}{cool}$ ($\Gamma$)           & 2.3    & 4 \\
$\ti{\delta}{rep}$ ($\Gamma$)            & 2.3    & 4 \\
$\ti{I}{cool}$ per beam ($\ti{I}{sat}$)   & 14    & 14 \\
$\ti{I}{cool}/\ti{I}{rep}$               & 8    & 20 \\
\colrule
D line properties & $^{40}$K & $^6$Li \\
\colrule
$\Gamma/(2\pi)$ (MHz)                    & 6.04  &5.87 \\
$\ti{I}{sat}$ (mW/cm$^2$)                & 1.75  &2.54 \\
\end{tabular}
\end{ruledtabular}
\caption{\label{tab:D1_Molasses_Parameters}Parameters of the simultaneous $^6$Li and $^{40}$K D$_1$ cooling phase.}
\end{minipage}
\end{table}
The presence of the other species reduces the atom numbers in the MOTs by $4\ze{\%}$ for $^{6}$Li and by $10\ze{\%}$ for $^{40}$K. However, we observe no mutual influence during the CMOT and the D$_1$ molasses phase. The temperatures and relative atom numbers in dual-species operation do not differ from single-species operation. This has several reasons. First, the D$_1$ resonances and lasers are ${\sim}100\ze{nm}$ apart in wavelength. Second, the CMOT and molasses phases are short in duration (5ms) and the light-induced interspecies collision losses or heating are minimized as atoms are accumulated in dark states.
\Cref{tab:Performance} summarizes the performance of the different experimental phases in dual species-operation.
For both $^{6}$Li and $^{40}$K the D$_1$ molasses phase largely reduces the temperature while the cloud-size after the CMOT phase is conserved. For both species this leads to a phase-space density close to $10^{-4}$.
%
\begin{table}[!htbp]
\begin{ruledtabular}
\renewcommand{\arraystretch}{1.2}
\begin{tabular}{ccccc}
  & T& N&  n & $\phi$\footnote{The given phase-space density does not take into account the different internal states and is calculated as $\phi=n\ti{\lambda}{B}^3$, where $\ti{\lambda}{B}$ is the thermal de Broglie wavelength.} \\
  & ($\ze{\mu K}$) & ($\times 10^9$) &  ($\times 10^{10}\ze{cm^{-3}}$) & ($\times 10^{-5}$)\\
  \colrule
  \multicolumn{5}{l}{\textsc{Lithium}}\\
  \colrule
  MOT & 1000 & 2 &  2.6& 0.03 \\
  CMOT & 800 & 2 & 18&  0.29 \\
  Molasses & 48 & 1.2 & 7.6&  8.2 \\
  \colrule
  \multicolumn{5}{l}{\textsc{Potassium}}\\
  \colrule
  MOT & 240 & 3.2 &  7& 0.02 \\
  CMOT & 2000 & 3.2 &  37& 0.06 \\
  Molasses & 11 & 3.2 &  30& 10.7 \\
\end{tabular}
\end{ruledtabular}
\caption{\label{tab:Performance}Performance of the different experimental phases for $^{6}$Li and $^{40}$K, in dual species operation. We show the optimum temperature $T$, the atom number $N$, the density $n$ and the phase-space density $\phi$.}
\end{table}

\section{\label{sec:Conclusion}Conclusion}

In this study we have investigated the properties of D$_1$ laser cooling both experimentally and with numerical simulations. The simulations take into account all relevant Zeeman and hyperfine levels as well as the three dimensional bichromatic lattice geometry. Simulations and experimental results match fairly well for both lithium and potassium. Various sub-Doppler cooling effects~\cite{Fernandes2012} are recovered in the full model.
We have outlined the importance of coherences between the ground-state hyperfine levels~\cite{Grier2013} and interpreted the cooling mechanisms as resulting from a combination of VSCPT-like non-adiabatic transitions  between dark and coupled states and Sisyphus-type cooling. The discrepancy (factor two to four) between the temperature predicted by the semi-classical  model and the experimentally observed ones calls for further investigations and the development of a full quantum treatment of the external atomic motion using a Monte-Carlo wave function approach \cite{Dalibard1992, Castin1995}.

We discussed the physics behind the robustness of the D$_1$ cooling scheme, in particular its insensitivity to the excited state hyperfine splitting and to the relative phase between the cooling and repumping lattices, which is in sharp contrast to its D$_2$ counterpart~\cite{Bambini1997, Xu2003, Gupta1993}. We first suggest and numerically verify that due to the predominance of the ``open transitions'' at D$_1$, the bichromatically dressed ground states exhibit a robust correlation between light-shift and decay. We clarify that such a correlation leads to accumulation of atomic population to the lowest energy dressed states at Raman-resonance for both non-adiabatic and Sisyphus cooling. The picture also helps to explain the enhanced cooling at Raman resonance, as well as the reduced cooling or even heating at large Raman detunings.
Because of a smaller absorption cross-section for atoms cooled in weakly coupled states, D$_1$ gray molasses should also be less affected by the density dependent heating than their D$_2$ counterparts~\cite{Drewsen1994}.

Experimentally, using commercial semiconductor laser sources delivering ${\sim}200\ze{mW}$ of CW power, we achieve efficient, simultaneous cooling of $^6$Li and $^{40}$K, resulting in a phase space density close to $10^{-4}$ for both species.
This D$_1$ cooling scheme enables efficient direct loading of a dipole or magnetic trap because of the large gain in temperature.
As recently shown in \cite{Burchianti2014a,Salomon2013} these conditions are well suited to directly load an optical dipole trap and to perform all-optical evaporation to quantum degeneracy. In our own experiments, we load a magnetic trap, transport the atoms to a separate science cell, and perform evaporative cooling of $^{40}$K in two Zeeman states with a combined magnetic/optical trap scheme introduced in \cite{Lin2009}. Deep quantum degeneracy ($T/T_F= 0.14)$ in the dipole trap has been achieved and will be the subject of a future publication.

Finally we have also used the D$_1$ gray molasses scheme to cool the bosonic $^{41}$K isotope. All of $5\times 10^9$ $^{41}$K atoms from a CMOT were cooled to a final temperature of $20\ze{\mu K}$ leading to a phase-space density of $1.1\times10^{-4}$. This confirms the generality of this D$_1$ sub-Doppler cooling method.

\FloatBarrier

\begin{acknowledgements}
We acknowledge useful discussions with A. T. Grier, I. Ferrier-Barbut, B. S. Rem, M. Delehaye, S. Laurent, and J.~V.~Porto, and support from R\'egion Ile de France (DIM Nano-K and IFRAF), EU (ERC
grants Ferlodim and Thermodynamix), Institut de France (Louis D. Award), and Institut Universitaire de France.
D.R.F. acknowledges the support of Funda\c{c}\~{a}o para a Ci\^{e}ncia e Tecnologia (FCT-Portugal), through the grant number SFRH/BD/68488/2010. S.W. acknowledges the support of the Physics Department at Swansea University when part of this research was carried out.
\end{acknowledgements}

\appendix

\section*{\label{sec:Experimental_Setup}Appendix: Experimental details}

In this section we describe the experimental details, as well as results of additional measurements on the D$_1$ molasses scheme, in particular, the single species operation of $^6$Li.

Our experimental setup has been already described previously~\cite{Ridinger2011}.
A Zeeman-slower for $^{6}$Li and a 2D$^+$-MOT for $^{40}$K load the three-dimensional dual-species MOT in the MOT-chamber.
The D$_2$ laser systems for $^{6}$Li and $^{40}$K comprise master oscillator power amplifiers (MOPAs) to produce light at 671\,nm and 767\,nm respectively. Beamsplitters and acousto-optic modulators (AOMs) generate the cooling and repumping beams, which are combined before injecting tapered amplifiers for the Zeeman-slower and 3D-MOT for $^{6}$Li and accordingly the 2D$^+$-MOT and 3D-MOT for $^{40}$K.

The D$_1$ laser system for $^{40}$K operates at 770\,nm and is composed of a MOPA and an electro-optic modulator (EOM) to produce the repumping frequency. The total power used for the $^{40}$K cooling is 240\,mW, with an intensity per molasses beam of $14\,{\ti{I}{sat}}$.

\begin{figure}[htbp]
\begin{center}
\includegraphics{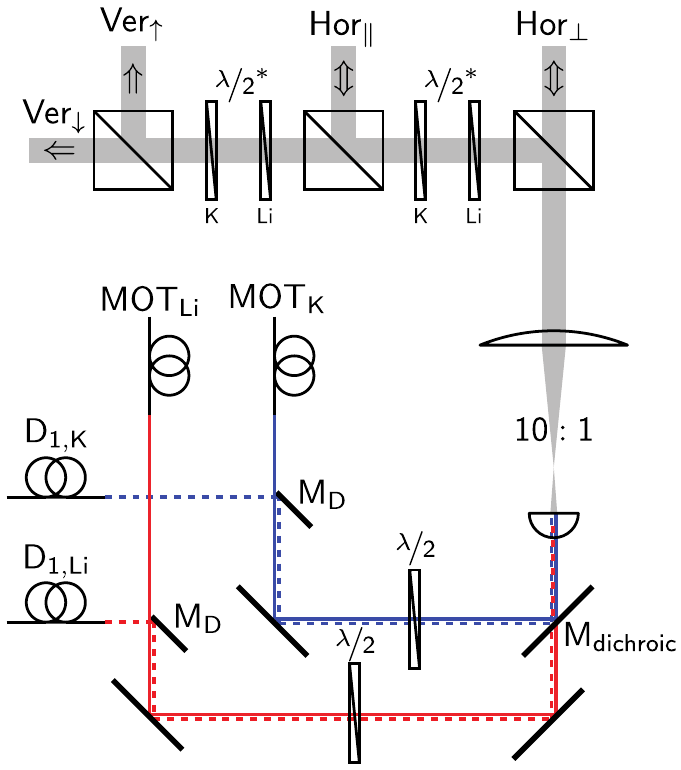}
\caption{\label{fig:Optical_scheme}(Color online) Optical scheme of the $D_1$-molasses. The 3D-MOT light and the D$_1$ cooling light are superposed using a D-shaped mirror ($\ti{M}{D}$). Afterwards a dichroic mirror ($\ti{M}{dichroic}$) combines the lithium- and potassium light, which is subsequently expanded and distributed to the three perpendicular axes of the 3D-MOT.}
\end{center}
\end{figure}

The source for the $^{6}$Li D$_1$ light at 671\,nm, used in this work, is a home-made solid-state laser, the next generation of ~\cite{Eismann2012,Eismann2013}, with up to $5\ze{W}$ output power.
AOMs allow to independently tune the frequencies and powers of the cooling and repumping beams, before recombination and injection into an optical fiber. We typically use $300\ze{mW}$ total power for the $^{6}$Li D$_1$ cooling. The waist of the $^6$Li D$_1$ beam after the telescope (\cref{fig:Optical_scheme}) is $8.6\ze{mm}$. We have also used a commercial 671\,nm tapered amplifier system (MOPA) with 130\,mW available power impinging on the atoms and obtained similar performances for the capture efficiency and sub-Doppler temperatures.

Our optical scheme superimposes the D$_1$ and D$_2$ light for both $^6$Li and $^{40}$K and produces the molasses and 3D-MOT beams (\cref{fig:Optical_scheme}). D-shaped mirrors ($\ti{M}{D}$) superpose the D$_1$ cooling light and the 3D-MOT light of each species before a dichroic mirror ($\ti{M}{dichroic}$) combines the lithium and potassium light. The beam containing all eight frequencies is expanded and distributed to the three pairs of $\sigma^+$--$\sigma^-$ counter-propagating beams of the 3D-MOT and the D$_1$ molasses. The two horizontal axes are retroreflected, the vertical axis consists of two independent beams.
The $\nicefrac{\lambda}{2}$ plates of order four for lithium ($\nicefrac{\lambda}{2}^{*}\tio{Li}$) and potassium ($\nicefrac{\lambda}{2}^{*}\tio{K}$) allow for independent control of the $^6$Li and $^{40}$K MOT power distribution.

The experiment starts with loading the dual-species MOT. In $10\ze{s}$ we typically load $8\times 10^8$ $^6$Li atoms with an initial temperature of $1\ze{mK}$ and $3\times 10^9$ $^{40}$K atoms at $200\ze{\mu K}$.
Then a CMOT phase~\cite{Mewes1999} increases the density of the atom cloud. The magnetic gradient is linearly ramped from $9\ze{G/cm}$ to $60\ze{G/cm}$ in $5\ze{ms}$. Meanwhile the frequencies of the cooling and the repumping beams are tuned closer to resonance and their intensities are linearly decreased. The CMOT phase results in an increase of the peak density by a factor of 7 (5.3) and a temperature of $800\ze{\mu K}$ ($2\ze{mK}$) for $^{6}$Li ($^{40}$K). At the end of the CMOT phase the current of the MOT-coils is switched off within ${\sim} 100\ze{\mu s}$.
We start the D$\tio{1}$ molasses phase for $^{6}$Li with a delay of $200\ze{\mu s}$ in order to wait for transient magnetic fields to decay. We found that this delay is not needed for $^{40}$K.

\subsection{$^{6}$Li D$_1$ molasses}

\begin{figure}
\begin{center}
\includegraphics{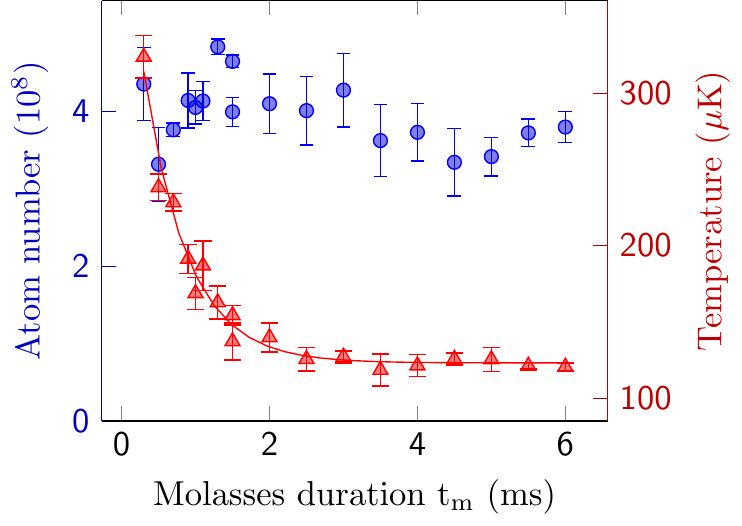}
\caption{\label{fig:Li_duration}(Color online) Number of atoms captured in the $^6$Li D$_1$ molasses (circles) and their temperature (triangles) as functions of molasses duration, with a $1/e$ cooling time constant $\ti{\tau}{cool}=0.6\ze{ms}$. The number of atoms in the compressed MOT is $8\times10^8$.}
\end{center}
\end{figure}

\begin{figure}
\begin{center}
\includegraphics{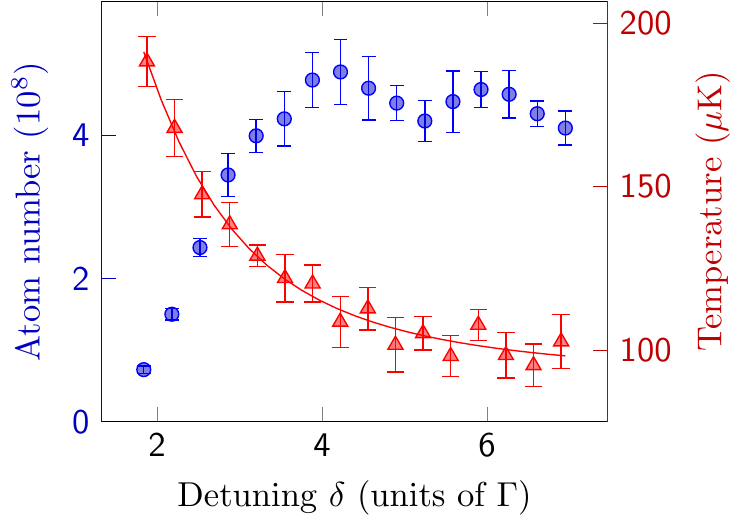}
\caption{\label{fig:Li_detuning}(Color online) Number of atoms captured in the $^6$Li D$_1$ molasses (circles) and their temperature (triangles) after a $3\ze{ms}$ capture phase at high intensity $\ti{I}{cool}=14.6 \,\ti{I}{sat}$ as functions of the global detuning $\delta=\ti{\delta}{cool}=\ti{\delta}{rep}$. The number of atoms in the compressed MOT is $8\times10^8$.}
\end{center}
\end{figure}

\begin{figure}[htbp]
\begin{center}
\includegraphics{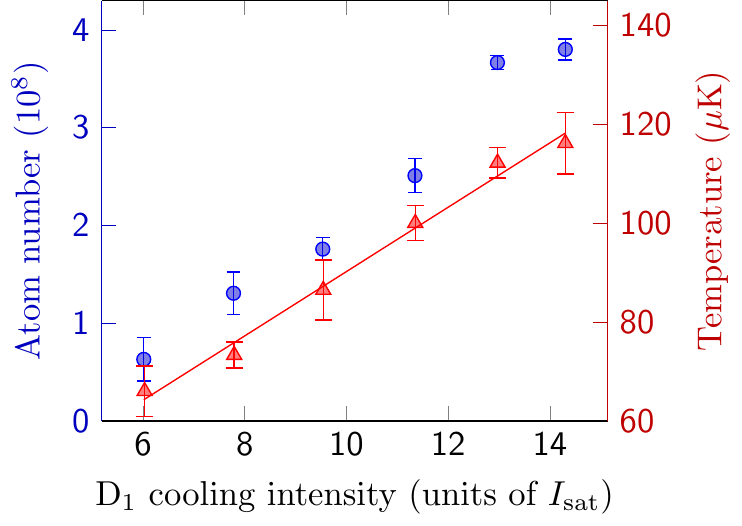}
\caption{\label{fig:Li_intensity}(Color online) Number of atoms captured in the $^6$Li D$_1$ molasses (circles) and their temperature (triangles) as functions of the D$_1$ cooling beam intensity for $\ti{\delta}{cool}=4\ze{\Gamma}$ and $\ti{I}{rep}=\ti{I}{cool}/20$. The number of atoms in the compressed MOT is $8\times10^8$. The atom number (temperature) increases linearly with a slope of $4\times10^7\ze{atoms/\ti{I}{sat}}$ ($6.5\ze{\mu K/\ti{I}{sat}}$).}
\end{center}
\end{figure}

\begin{figure}[htbp]
\begin{center}
\includegraphics{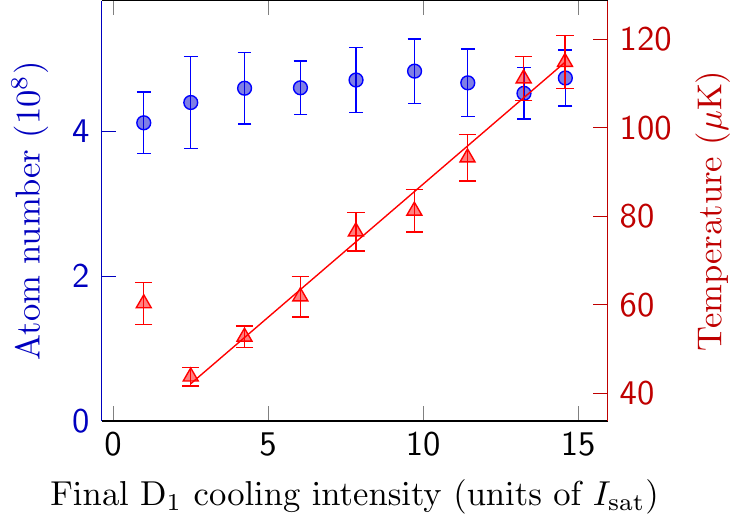}
\caption{\label{fig:Li_intensity_ramp}(Color online) Number of atoms captured in the $^6$Li D$_1$ molasses (circles) and their temperature (triangles) after a $3\ze{ms}$ capture phase at high intensity $\ti{I}{cool}=14.6 \ti{I}{sat}$ followed by a 2ms linear intensity ramp to an adjustable value. The temperature increases linearly for larger intensity with a slope of ${\sim}6\ze{\mu K/\ti{I}{sat}}$ The detuning is fixed to $\ti{\delta}{cool}=4\ze{\Gamma}$. The number of atoms in the compressed MOT is $8\times10^8$.}
\end{center}
\end{figure}

Here we study the cooling dynamics of $^6$Li alone. We set the peak intensity of the molasses to 14.6$\,{\ti{I}{sat}}$ per beam, the cooling/repumping ratio to $\ti{I}{cool}/\ti{I}{rep}=20$ and fix the global detuning to $\delta=\ti{\delta}{cool}=\ti{\delta}{rep}=4\ze{\Gamma}$. Here $\ti{I}{sat}=2.54\ze{mW/cm^2}$ is the saturation intensity of the D$_2$ cycling transition and $\Gamma=2\pi\times5.87\ze{MHz}$ the D$_1$ line natural linewidth for $^6$Li. \Cref{fig:Li_duration} shows the atom number and temperature of the D$_1$ molasses as functions of the molasses duration $\ti{t}{m}$. The temperature is determined through time of flight measurements. We capture $60\ze{\%}$ of the $8\times10^8$ CMOT atoms. Within $3\ze{ms}$ the atoms are cooled from $800\ze{\mu K}$ to the asymptotic temperature of $120\ze{\mu K}$ with a $1/e$ cooling time constant $\ti{\tau}{cool}=0.6\ze{ms}$. The direct measurement of the fluorescence shows a very fast decay at the beginning ($\ti{\tau}{fast}<1\ze{\mu s}$), followed by a further decrease by another factor of two within $2\ze{ms}$. This indicates that the atoms are accumulated in dark states during the cooling process. We find the 1/e-lifetime of the D$_1$ molasses atom number for these parameters to be $\ti{\tau}{D1}=90\ze{ms}$.
The molasses atom number and temperature as functions of the global detuning $\delta$ is shown in \cref{fig:Li_detuning}. We observe a decrease of the temperature from $188\ze{\mu K}$ to $100\ze{\mu K}$ for $\ti{\delta}{cool}=2\dots7$. The capture efficiency raises sharply until $4\ze{\Gamma}$ and stays constant until $7\ze{\Gamma}$.
We now study the influence of the D$_1$ light intensity. When increasing the cooling intensity, while keeping the molasses time fixed to $3\ze{ms}$, we observe both an increase of the capture efficiency and of the temperature (\cref{fig:Li_intensity}).
To take advantage of the intensity dependent temperature, we use two successive phases in the cooling sequence. During the first $3\ze{ms}$ we capture the atoms at high intensity, yielding the highest capture efficiency. Then the intensity is linearly ramped within $2\ze{ms}$ to an adjustable final intensity to further lower the temperature. \Cref{fig:Li_intensity_ramp} shows that the intensity ramp reduces the final temperature from $115\ze{\mu K}$ to $44\ze{\mu K}$ without a significant atom loss for a final intensity $\ti{I}{cool,final}=2.5\ti{I}{sat}$. For lower intensities we observe heating and atom loss.

The cooling/repumping intensity-ratio influences the atom number and the temperature. We find an optimal temperature for $\ti{I}{cool}/\ti{I}{rep}=20$. For a smaller ratio the temperature increases slightly and the atom number starts to drop at around $\ti{I}{cool}/\ti{I}{rep}=7$. For $\ti{I}{cool}/\ti{I}{rep}>33$ the cooling mechanism becomes inefficient, leading to atom loss and heating.\vspace{0.0cm}

Measuring the atom cloud size at the end of the CMOT and molasses phases we see no significant change, proving that the diffusion during the molasses phase is small. In terms of phase-space density the atom loss is largely overcompensated by the 14 times reduction in temperature.
\FloatBarrier

%

\end{document}